\documentclass[prb, epsf]{revtex4}
\usepackage{graphicx}
\usepackage{dcolumn}
\usepackage{bm}
\usepackage[T2A]{fontenc}
\usepackage[cp1251]{inputenc}
\usepackage[russian]{babel}
\usepackage{bm}
\begin{document}
\title{Direct numerical observation of Euclidean resonance}
\author{V.V. Makhro}
\email{Makhro@mail.ru} \affiliation{Irkutsk State University,
Bratsk Branch, 665708, Bratsk, prospekt Lenina, 34}

\date{August 20, 2004 }

\begin{abstract}
We present results of direct numerical calculations for the problem of quantum tunneling through the time-dependent potential barrier. Computations clearly demonstrate existence of the effect of the underbarrier resonance. 

Мы представляем результаты прямого численного решения задачи квантового туннелирования сквозь потенциальный барьер, зависящий от времени. Вычисления отчетливо демонстрируют существование эффекта подбарьерного (эвклидова) резонанса.

\end{abstract}

\maketitle

The dynamics of tunneling processes has been under discussion for a long time. For obvious reasons, the problem how to increase tunneling rate in different situations has especially interest. A spell ago, we investigate some aspects of this problem in our paper \cite{ftt98}, \cite{ssp98}. In particular, we found two interesting mechanisms for stimulation of tunneling. 

First one is so-called ``thermal stimulation''  when the tunneling particle absorb some quanta of energy in front of barrier, lifts to top of barrier and then tunnel in more transparent region of barrier (thermal assisted or photon assisted tunneling). 

The second way to increase tunneling rate that we discovered is the underbarrier resonant stimulation, which some authors named later as ``Euclidean resonance'' \cite{ivlev04}. The main idea is as follows.  In imaginary time underbarrier motion of particle one can consider as a particle motion in some potential well (overturned barrier). We show that external perturbation with an appropriate frequency could pull down the particle to the bottom of potential well (top of the barrier in real time) and essentially increase tunneling rate as a whole. Very important is the fact that harmonic perturbation in imaginary time must be in form of decreasing exponent in real time. 

Now we find evidence of this fact by the direct numerical solving of Schr\"{o}dinger equation for the particle in the field of time-dependent potential barrier.

We consider a simplest case for the tunneling through the time-dependent rectangular barrier. In our calculations we use an implicit numerical scheme for the 1-d Schr\"{o}dinger equation

\begin{equation}
i\hbar \frac{{\partial u}}
{{\partial t}} =  - \frac{{\hbar ^2 }}
{{2m}}\frac{{\partial ^2 u}}
{{\partial x^2 }} + V\left( {x,t} \right)u
\end{equation}

which has the form

\begin{equation}
u_{i,j}  = Eu_{i,j + 1}  - Fu_{i + 1,j + 1}  - Fu_{i - 1,j + 1}
\end{equation}

where we put $E = \frac{1}{K}\left( {1 + 2\mu } \right)$, $F = \frac{1}
{K}\mu $, $K = 1 - Dt$, $D = \frac{i}{\hbar }V\left( {x,t} \right)$ and $\mu  = \frac{{i\hbar }}{{2m}}\frac{t}{{h^2 }}$. $t$ is the temporal step of the difference scheme and $h$ is the spatial one. 

\begin{figure}[hbt]
\includegraphics[width=0.5\textwidth]{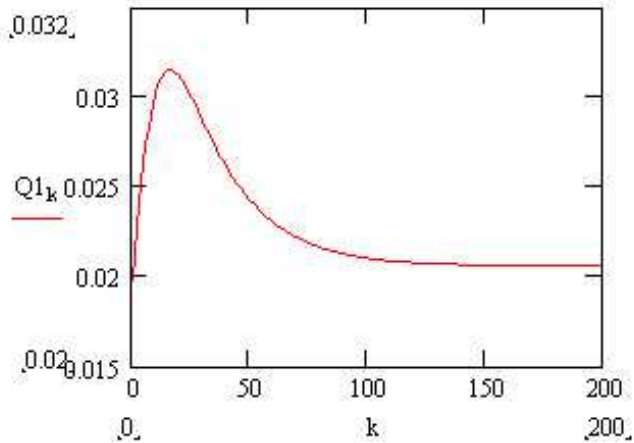}
 \caption{Probability $Q$ of the underbarrier resonant tunneling as the function of the imaginary frequency $k$}
  \label{Fig1}
\end{figure}

For simplicity, we took dimensionless parameters and put $\hbar =1$. 
Below we consider for an example some situation, which is in one sense a good model for mesoscopic problems. 
Thus, the initial condition was 
	
\begin{equation}
u_{i,1}  = \beta \cos \left( {f\pi \frac{i}
{n}} \right)
\end{equation}
for $i \in \left[ {1,\frac{n}{3}} \right]$ and boundary conditions were

\begin{equation}
u_{1,j}  = 0,u_{n,j}  = 0
\end{equation}	
for $j \in \left[ {1,m} \right]$. $i$ is current index of spatial step, $j$ is temporal one, $n$ -- number of spatial steps and $m$ - number of temporal steps. The barrier function we define as
	
\begin{equation}
v_{i,j}  = \left( {1 + \varepsilon \exp \left( { - \frac{j}
{m}k} \right)} \right)
\end{equation}
for $i \in \left[ {n/3,n/3 + \delta } \right]$ and $v_{i,j} = 0$ for $i < n/3$ and $i > n/3 + \delta$. $k$ is a parameter for the trimming of damping constant in exponent (or resonant frequency in imaginary time), $\delta$ is the width of barrier. For each parameter $k$ we calculate tunneling probability $Q$ as the ratio of integral of $\left| {u\left( {x,t} \right)} \right|^2 $ over ante-barrier region to some one over space behind the barrier. 

Results of calculations are shown in Fig. 1 where we present dependence of tunneling probability $Q$ vs $k$ for the initial parameters $F=-1.786i$, $E=1+3.571i$ outside the barrier and $F=0.171-1.769i$, $E=0.648+3.634i$ in the underbarrier region. 

It is easy to see, that this dependence really has resonant character. We intend publish more detailed results soon but now we believe that founded features is a good confirmation for the Euclidean resonance.

In Appendix we place the code of our computing programme.
\begin{appendix}
\begin{figure}[hbt]
\includegraphics[width=0.5\textwidth]{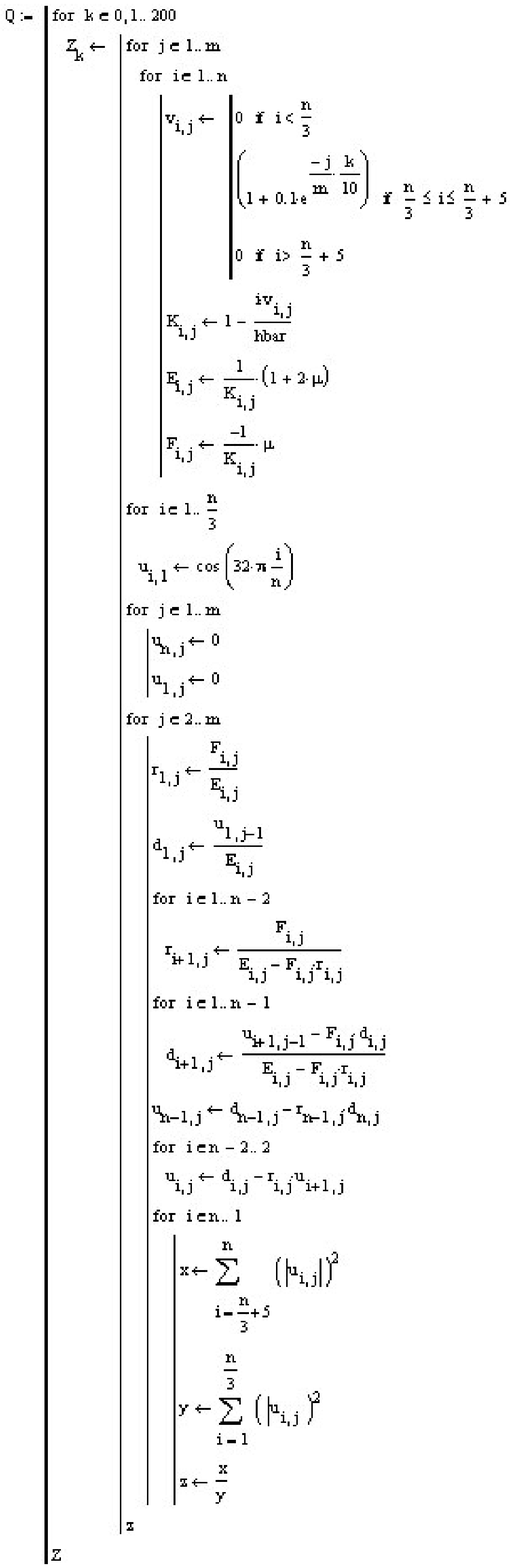}
 \caption{The Mathcad code}
  \label{Fig1}
\end{figure}

\end{appendix}


\begin{thebibliography}{10}
\bibitem{ftt98}
В. В. Махро, ФТТ {\bf 40}, 1855  (1998).
\bibitem{ssp98}
V. V. Makhro, Physics of the Solid State, {\bf 40}, 1681  (1998).
\bibitem{ivlev04}
B. Ivlev, quant-ph/0407163 v1 (2004).

\end{thebibliography}
\end{document}